\documentclass[a4paper,12pt]{article}
\usepackage{amsmath}
\usepackage{graphicx}
\usepackage{indentfirst}
\usepackage{setspace}

\title{Relative Resolution: \\ A Hybrid Formalism for Fluid Mixtures}
\author{Aviel Chaimovich \\ Christine Peter and Kurt Kremer}
\date{}

\begin{document}

\pagestyle{plain}
\onehalfspacing
\maketitle

\clearpage
\begin{abstract}
We show here that molecular resolution is inherently hybrid in terms of relative separation:  
If molecules are close to each other, they must be characterized by a fine-grained (geometrically detailed) model, yet if molecules are far from each other, they may be described by a coarse-grained (isotropically simplified) model.  
We notably present an analytical expression for relating the two models by energy conservation.  
This hybrid framework is correspondingly capable of retrieving the structural and thermal behavior of various multi-component and multi-phase fluids across state space.  
\end{abstract}

\clearpage
\section{Introduction}

\subsection{}

Ever since the days of van der Waals, expressing molecular interactions in terms of short-range and long-range contributions has proven insightful.  

In such a spirit for uniform (nonpolar) liquids, ``force truncation'' recognizes that if molecules are close to each other, their forces must be entirely considered, yet if molecules are far from each other, their forces may be effectively smeared \cite{Frisch_ACP1964,Widom_Science1967}.  

This is because structural correlations are dominated by the fluctuating energetics of nearest neighbors, and thermal properties are corrected for by the mean field of other neighbors.  

Importantly, while the Weeks-Chandler-Anderson (WCA) study has notably established to truncate the force between its repulsive and attractive portions \cite{ChandlerWeeks_PRL1970,WeeksAndersen_JCP1971}, the Ingebrigtsen-Schroder-Dyre (ISD) study has recently argued to truncate it between its primary and secondary coordination shells \cite{ToxvaerdDyre_JCP2011,IngebrigtsenDyre_PRX2012}.  

Nevertheless for multi-component or multi-phase systems, long-range interactions markedly influence segregation, as short-range interactions merely affect packing, and thus, ``force truncation'' cannot generally describe such nonuniform scenarios \cite{AndersenWeeks_ACP1976,BarkerHenderson_RMP1976}.  

While an elegant extension has been developed, it is only capable of describing idealized nonuniformities \cite{WeeksBroughton_PRL1995}.

\subsection{}

In fact, nonuniform liquids have received much attention by an apparently unrelated community which connects fine-grained (FG) and coarse-grained (CG) models in various ways.  

In particular, Adaptive Resolution contains both FG and CG models in a single system, switching between them in terms of absolute position (e.g., a molecule is represented by a FG potential if close to the origin and by a CG potential if far from the origin) \cite{PraprotnikKremer_JCP2005,Abrams_JCP2005,EnsingParrinello_JCTC2007,PotestioDonadio_PRL2013}.  

While such hybrid strategies appear to be computationally useful in examining mixtures, the greatest challenge in these approaches is the relationship between the FG and CG models.  

While there are numerous protocols for yielding a CG model given a FG model, there is a general consensus between these numerical parametrizations:  
No strategy optimally transfers in state space while concurrently representing the structural and thermal behavior of the reference mixture \cite{Louis_JPCM2002,RuhleAndrienko_JCTC2009,PeterKremer_FD2010,Noid_JCP2013}.

\subsection{}

In our current work, we present Relative Resolution (RelRes):  
Molecules interact with each other via a FG modeling \emph{resolution} at small \emph{relative} separation and via a CG modeling \emph{resolution} at large \emph{relative} separation.  

While a reminiscent idea has been recently motivated by Adaptive Resolution as well \cite{ShenHu_JCTC2014}, we here essentially formalize RelRes as an extension of ``force truncation'' beyond uniform liquids; rather than arbitrarily linking the models, we here impose energy conservation, in turn formulating an analytical relationship between the FG and CG potentials.  

This parametrization enables RelRes to capture across state space the structural and thermal behavior of various mixtures, and we consequently present a hybrid description for multi-component and multi-phase systems:  
If molecules are close to each other, their geometrical intricacies are crucial for consideration, yet it they are far from each other, they may be characterized as isotropic points.

\clearpage 
\section{Theoretical Foundation}

\subsection{}

Before describing RelRes, we formulate nonhybrid systems.  

A CG mixture is frequently composed of single-site molecules, governed by the following Hamiltonian:

\begin{equation}
U^{CG} = \frac{1}{2} \sum_{{i}\neq{j}} u^{CG}_{ij} \left( r_{ij} \right)
\label{eqn:PureCG}
\end{equation}

Corresponding usually with gravitational centers, a particular pair $ij$ interacts via its relative separation $r_{ij}$ by a CG potential $u^{CG}$.  

A FG mixture is normally composed of multi-site molecules, governed by the following Hamiltonian:

\begin{equation}
U^{FG} = \frac{1}{2} \sum_{{i}\neq{j}} \sum_{\alpha_i\alpha_j} u^{FG}_{\alpha_i\alpha_j} \left( r_{ij}+\Delta{r}_{\alpha_i\alpha_j} \right)
\label{eqn:PureFG}
\end{equation} 

Corresponding typically with atomistic coordinates, a particular pair $\alpha_i\alpha_j$ interacts via its relative separation $r_{ij}+\Delta{r}_{\alpha_i\alpha_j}$ by a FG potential $u^{FG}$.  

In our entire formalism, we ignore intramolecular energetics; our hybrid approach maintains all degrees of freedom, and therefore, intramolecular forces remain unaltered.

\subsection{}

RelRes combines both of the above in the following way:  

\begin{equation}
\tilde{U} = \frac{1}{2} \sum_{{i}\neq{j}} \left[ \sum_{\alpha_i\alpha_j} \tilde{u}^{FG}_{\alpha_i\alpha_j} \left( r_{ij}+\Delta{r}_{\alpha_i\alpha_j} \right) + \tilde{u}^{CG}_{ij} \left( r_{ij} \right) \right]
\label{eqn:RelRes}
\end{equation}

The focal idea here is that while $\tilde{u}^{FG}$ is the short-range portion of the FG potential, $\tilde{u}^{CG}$ is the long-range portion of the CG potential, and we particularly switch between these respectively in the following way:  

\begin{equation}
\tilde{u}^{FG}_{\alpha_i\alpha_j} \left( r \right) =  
\begin{cases}
u^{FG}_{\alpha_i\alpha_j} \left( r \right) - u^{FG}_{\alpha_i\alpha_j} \left( r_s \right) & \text{if  } r \leq r_s
\\
0 & \text{if  } r \geq r_s
\end{cases}
\label{eqn:HybridFG}
\end{equation}

\begin{equation}
\tilde{u}^{CG}_{ij} \left( r \right) = 
\begin{cases}
u^{CG}_{ij} \left( r_s \right) & \text{if  } r \leq r_s
\\
u^{CG}_{ij} \left( r \right) & \text{if  } r \geq r_s
\end{cases}
\label{eqn:HybridCG}
\end{equation}

$r_s$ is the distance at which the information is switched.  

Note that $U^{FG}=U\left(r_s\to\infty\right)$ and $U^{CG}=U\left(r_s\to0\right)$.

\subsection{}

We now seek a relationship between the FG and CG potentials.  

RelRes naturally accounts for the FG interactions at small separations, and thus, we solely require a parametrization for the CG interactions at large separations.  

Consequently, we consider the appropriate limit of the nonhybrid systems, as defined by Eqn.~\ref{eqn:PureCG} and Eqn.~\ref{eqn:PureFG}, in turn demanding energy conservation between them; specifically, we equate these equations, while assuming that each system is merely composed of two molecules whose relative distance approaches infinity:  

\begin{equation*}
u^{CG}_{ij} \left( r_{ij} \right) = \lim_{r_{ij}\to\infty} \sum_{\alpha_i\alpha_j} u^{FG}_{\alpha_i\alpha_j} \left( r_{ij}+\Delta{r}_{\alpha_i\alpha_j} \right)
\end{equation*}
\begin{equation}
u^{CG}_{ij} \left( r \right) = \sum_{\alpha_i\alpha_j} u^{FG}_{\alpha_i\alpha_j} \left( r \right)
\label{eqn:BaseMPIL}
\end{equation}

This equation analytically parametrizes from a FG model to a CG model, and for the remainder of our work, we term it as \emph{the Molecular Pair in the Infinite Limit} (MPIL).  

MPIL is alternatively retrieved by vanishing molecular dimensions in a multipole expansion whose monopole term is nonzero; with the FG atomistic coordinates collapsing onto the CG gravitational centers, MPIL geometrically translates as the summation of all interactions as if they are interconnecting only two isotropic points.

\subsection{}

Suppose that the FG and CG potentials can be each cast as a linear combination of basis functions, $ {u}^{FG}_{\alpha_i\alpha_j} \left( r \right) = \sum_l c^{FG}_{l,\alpha_i\alpha_j} f^{FG}_{l} \left( r \right) $ and $ {u}^{CG}_{ij} \left( r \right) = \sum_l c^{CG}_{l,ij} f^{CG}_{l} \left( r \right) $, respectively.  
  
Here, $l$ is the index of the basis function $f_l$, and $c_l$ is its respective coefficient.  

Substituting these expressions in Eqn.~\ref{eqn:BaseMPIL}, we make the following observations:

\begin{equation}
f^{CG}_{l} \left( r \right) = f^{FG}_{l} \left( r \right)
\label{eqn:FunctionMPIL}
\end{equation}

\begin{equation}
c^{CG}_{l,ij} = \sum_{\alpha_i\alpha_j} c^{FG}_{l,\alpha_i\alpha_j}
\label{eqn:CoefficientMPIL}
\end{equation}

These expressions respectively mean that the basis functions of the CG model must be the same as those of the FG model, and that the CG coefficients can be sequentially determined by elementary arithmetics of the FG coefficients.  

As a rudimentary example, consider the Lennard-Jones (LJ) potential for a system of identical molecules, each having $n$ identical sites.  

Not only that MPIL imposes LJ functionality for all interactions, but it also relates between the length and energy parameters, $\sigma^{CG}=\sigma^{FG}$ and $\epsilon^{CG}=n^2\epsilon^{FG}$, respectively.  

\begin{figure}[p]
\centering
\includegraphics{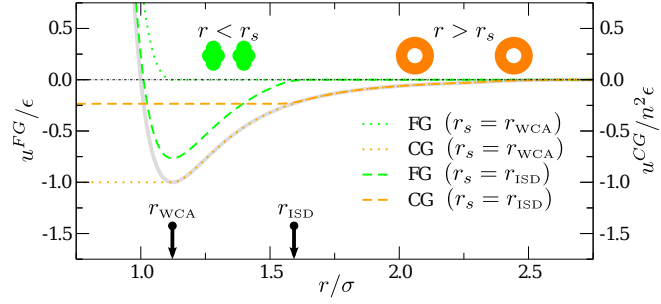}
\caption
{ 
Variations of FG (green curves, left ordinate) or CG (orange curves, right ordinate) potentials in terms of a relative separation.  
We nondimensionalize by the parameters of the former model throughout.  
The solid (gray) curve is for the full LJ potential.  
}
\label{Fig:HybridModel}
\end{figure}

In Fig.~\ref{Fig:HybridModel}, we exemplify RelRes with MPIL for this scenario, particularly having two variations for $r_s$:  
One is motivated by Ref.~\cite{ChandlerWeeks_PRL1970} (i.e., $r_{\text{\tiny{WCA}}}\equiv2^{1/6}\sigma$, dotted curves), while another is inspired by Ref.~\cite{ToxvaerdDyre_JCP2011} (i.e., $r_{\text{\tiny{ISD}}}\equiv2^{2/3}\sigma$, dashed curves).  

We continue examining these alternatives for $r_s$ throughout, importantly analyzing their influence on RelRes in describing \emph{reference mixtures} (i.e., systems with just FG and no CG interactions all towards infinity).

\clearpage
\section{Computational Validation}

\subsection{}

We proceed by testing these ideas with molecular simulations \cite{HessLindahl_JCTC2008}.  

Employing the leap-frog algorithm, we maintain a density $\rho$ and a temperature $T$ via canonical sampling \cite{BussiParrinello_JCP2007}.  

\begin{figure}[p]
\centering
\includegraphics{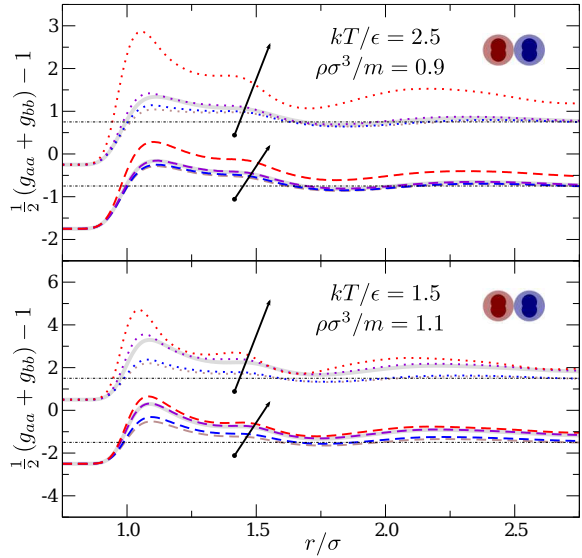}
\caption
{
Radial distributions for the ethane-like scenario in terms of the relevant distance.  
In each panel, the top (dotted) set is for $r_s=r_{\text{\tiny{WCA}}}$, and the bottom (dashed) set is for $r_s=r_{\text{\tiny{ISD}}}$.  
The gray (solid) curve is for the reference mixture, and it is duplicated between the sets; this coding is maintained throughout our work.  
Arrows point in the direction of increasing $\lambda$.  
Each function is vertically shifted by an arbitrary constant, and its zero value is marked by its (adjacent) dotted-dashed line.  
}
\label{Fig:EthaneStructure}
\end{figure}

For a preliminary examination, we construct a binary ethane-like mixture (i.e., $n=2$ for both types), depicted in Fig.~\ref{Fig:EthaneStructure}.  

Importantly, CG $\mathcal{A}$ and $\mathcal{B}$ sites are strictly composed of FG $a$ and $b$ sites.  

We purposefully set the conditions of the reference mixture so as to yield a segregation between two liquids in ample time.  

While invoking Lorentz-Berthelot rules for mixing, the LJ potential is the basis for all interactions.  

Importantly, $\sigma^{CG}_{ij}=\sigma^{FG}_{\alpha_i\alpha_j}$ and $\epsilon^{CG}_{ij}=\lambda\epsilon^{FG}_{\alpha_i\alpha_j}$, with $\lambda$ being the tuning parameter which allows us to systematically examine our hybrid approach (i.e., $\lambda=0$ expresses ``force truncation'', and $\lambda=4$ retrieves MPIL parametrization).  

Finally, the mass of each FG site is $m$, and the mass of each CG site is zero; we represent Boltzmann's constant by $k$, giving further information in the caption of Fig.~1 of the SM [21].

\subsection{}

We focus foremost on the ability of the hybrid systems in describing structural correlations, as manifested by radial distributions, $g$.  

Technically partitioned in bins of size $\sigma/800$, we present various combinations of these functions in Fig.~\ref{Fig:EthaneStructure} here, as well as in Fig.~1 of the SM [21].  

Clearly in either panel, the traditional variation of ``force truncation'', as manifested by $\lambda=0$ with $r_s=r_{\text{\tiny{WCA}}}$ (i.e., dotted brown curve), is completely inadequate in describing the various $g$, particularly lacking any evidence of a phase segregation.  

Having only short-range repulsions, the various sites appear almost identical between each other at separations closer than $r_{\text{\tiny{WCA}}}$, and as such, they have a similar preference for each other in terms of structure.  

Keeping $\lambda$ constant while lengthening $r_s=r_{\text{\tiny{ISD}}}$ (i.e., dashed brown curve), we still cannot attain a sufficient representation of the phase segregation, suggesting in turn that energetics beyond $r_s$ may be necessary.  

Chiefly in either panel, the parametrization of MPIL, as exemplified by $\lambda=4$ with $r_s=r_{\text{\tiny{ISD}}}$ (i.e., dashed violet curve), is excellently capable of retrieving the various $g$, especially capturing all signatures of the phase segregation.  

It appears that the CG sites effectively correct for the FG sites in their long-range attractions, advocating that at separations beyond $r_{\text{\tiny{ISD}}}$, an isotropic assumption for molecules sufficiently describes the relevant forces.  

Keeping $\lambda$ constant while shortening $r_s=r_{\text{\tiny{WCA}}}$ (i.e., dotted violet curve), we cannot anymore attain a flawless representation of radial distributions, suggesting in turn that geometrical intricacies of all nearest neighbors may be vital for structure.  

Furthermore, MPIL parametrization is validated as we perturb it off its ideal value:  
For either value of $r_s$, the hybrid system of RelRes is under-segregated for $\lambda=2$ (i.e., blue curves) and over-segregated for $\lambda=6$ (i.e., red curves), as compared with the reference mixture.  

Although CG models cannot in general replicate the structure of FG models, RelRes does not have such issues as it makes a valid approximation beyond $r_s$ via MPIL.

\subsection{}

\begin{figure}[p]
\centering
\includegraphics{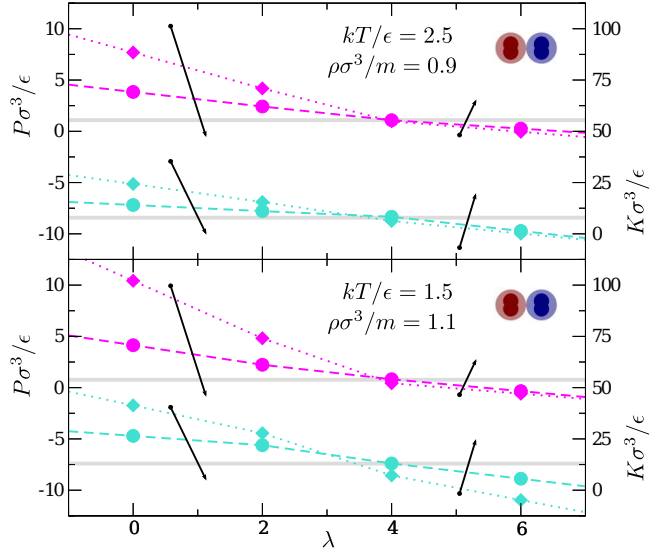}
\caption
{
Sets of pressure (magenta symbols, left ordinate) and modulus (turquoise symbols, right ordinate) in the ethane-like scenario in terms of the tuning parameter.  
Arrows point in the direction of increasing $r_s$.  
}
\label{Fig:EthaneEnergy}
\end{figure}

We now correspondingly examine thermal properties in Fig.~\ref{Fig:EthaneEnergy}:  
By the virial approach, we compute the pressure, $P$, and by inspecting adjacent densities, we also evaluate the modulus, $K$ (i.e., the inverse of the compressibility).  

Importantly, these two cannot be concurrently retrieved for a FG system by a CG system, but this is not the case for RelRes.  

For either $r_s$, both decrease almost linearly with increasing $\lambda$, and analogously with the observations above, each set crosses its reference value at a single point (i.e., $\lambda$ of MPIL).  

RelRes naturally maintains the energetics between near neighbors, and it also appears that via MPIL, it correctly describes the energetics between far neighbors; consequently, our hybrid approach can retrieve these thermal properties of the reference mixture.  

We also note here that the sets of $r_s=r_{\text{\tiny{ISD}}}$ (i.e., circles) consistently yield better replication than the sets of $r_s=r_{\text{\tiny{WCA}}}$ (i.e., diamonds).

\subsection{}

Importantly, our ethane-like analysis involves two different sets of temperature and density, as manifested by the panels of both Fig.~\ref{Fig:EthaneStructure} and Fig.~\ref{Fig:EthaneEnergy}.   

As such, it appears that RelRes with MPIL is transferable across $kT/\epsilon$ and $\rho\sigma^3/m$ of order unity, sufficiently representing the structural and thermal behavior of the reference mixture in this region.  

While CG models cannot be optimally transferable for FG models throughout state space, RelRes prevails such issues via MPIL:  

Transferability may stem in the fact that Eqn.~\ref{eqn:BaseMPIL} contains no state parameters; perhaps, state functionality is involved in $r_s$, but we do not observe such subtleties here.  

Regardless, we must emphasize that RelRes with MPIL is only necessary for mixtures; for uniform liquids, ``force truncation'' is usually sufficient, and we exemplify this via another ethane-like study which is described in Fig.~2 of the SM [21].

\subsection{}

Nevertheless, can RelRes with MPIL successfully describe realistic mixtures as well?  

\begin{figure}[p]
\centering
\includegraphics{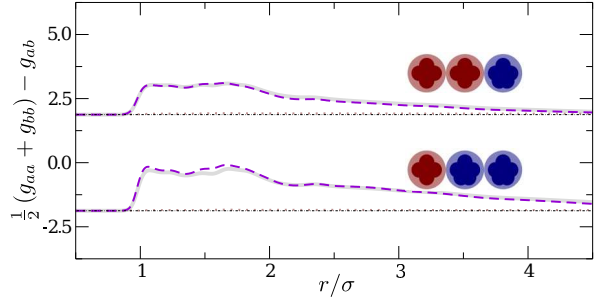}
\caption
{
Radial distributions for the pentane-like scenario in terms of the relevant distance.  
While these curves are in the spirit of Fig.~\ref{Fig:EthaneStructure}, the two sets here are for two compositions, and in each case, $kT/\epsilon=2.0$ and $\rho\sigma^3/m=1.0$.  
For the top scenario, the reference mixture has $P\sigma^3/\epsilon=-0.44$ and $K\sigma^3/\epsilon=-11.1$; these are respectively $\{-0.45,-11.5\}$ for RelRes with MPIL and $\{+3.38,+11.4\}$ for ``force truncation''.  
For the bottom scenario, the reference mixture has $P\sigma^3/\epsilon=-0.20$ and $K\sigma^3/\epsilon=-3.6$; these are respectively $\{-0.20,-3.2\}$ for RelRes with MPIL and $\{+2.90,+9.5\}$ for ``force truncation''.  
}
\label{Fig:PentaneStructure}
\end{figure}

We demonstrate that this is in fact the case by devising a binary pentane-like system (i.e., $n=5$ for both types), depicted in Fig.~\ref{Fig:PentaneStructure}.  

Importantly, this is generically representative of a tetrachloromethane-thiophene mixture. 

We purposefully set the conditions of the reference mixture so as to yield an instability via a cavity.  

We specifically focus on RelRes with MPIL, $\epsilon^{CG}_{\mathcal{AA}}=16\epsilon^{FG}_{aa}+8\epsilon^{FG}_{ab}+\epsilon^{FG}_{bb}$ and $\epsilon^{CG}_{\mathcal{BB}}=\epsilon^{FG}_{aa}+8\epsilon^{FG}_{ab}+16\epsilon^{FG}_{bb}$, comparing it in turn with ``force truncation''.  

We proceed in a reminiscent manner as with the ethane-like scenario, giving further information in the caption of Fig.~3 of the SM [21].

\subsection{}

We present various combinations of radial distributions in Fig.~\ref{Fig:PentaneStructure} here, as well as in Fig.~3 of the SM [21]; the caption of the former also correspondingly lists values for pressure and modulus.  

Above all, RelRes with MPIL at $r_s=r_{\text{\tiny{ISD}}}$ describes the structural and thermal behavior of the reference fluid with sufficient agreement, and this is especially notable as compared with ``force truncation'' at $r_s=r_{\text{\tiny{WCA}}}$,  which mistakenly predicts a stable mixture with no cavity.  

With our hybrid approach, molecules can readily distinguish between a liquidus droplet and a gaseous bubble at a certain distance, capturing in turn the appropriate energetics of their near and far neighbors.  

While this argues for the transferability across both gaseous and liquidus phases, RelRes with MPIL is also transferable across different compositions, since we vary the concentration in this tetrachloromethane-thiophene scenario.  

Finally, we construct another pentane-like scenario which is described in Fig.~4 of the SM [21]; we consequently show that our approach is capable of describing two liquids in coexistence with a gas, reiterating the robustness of RelRes with MPIL.

\clearpage
\section{Conclusion}

\subsection{}

In summary, our current work presents a hybrid picture for fluid mixtures:  
If molecules are close to each other, their geometrical intricacies are crucial for consideration, yet if they are far from each other, they may be characterized as isotropic points.  

We have demonstrated this idea by introducing here two distinct concepts, the framework of RelRes and the parametrization of MPIL, which are ideal if employed concurrently.  

RelRes sets the general Hamiltonian for switching between the FG and CG resolution in terms of relative separation between molecules, while MPIL provides the relationship between these two models (i.e., Eqn.~\ref{eqn:BaseMPIL}).  

As such, RelRes with MPIL is capable of retrieving across state space not only the structural correlations but also the thermal properties (e.g., pressure, compressibility, etc.) of the multi-component and multi-phase fluids which we have investigated here (e.g., a tetrachloromethane-thiophene mixture).  

Importantly, it appears that the energy conservation of MPIL is responsible for reproduction of thermal properties, since an approach reminiscent of RelRes only replicates structural correlations \cite{ShenHu_JCTC2014}.  

Besides, it is rather remarkable that the infinite limit involved in our formalism is virtually satisfied just beyond nearest neighbors.  

On a practical level, our hybrid approach presents a useful tool in exploring the behavior of mixtures.  

Except that RelRes notably diminishes the number of interactions necessary for computation beyond $r_s$, the analytical parametrization of MPIL allows for trivial transferability across state space.  

As a next step, it may be of value to incorporate some of these ideas in Adaptive Resolution.

\clearpage
\section*{Acknowledgments}
We are foremost grateful of the Alexander von Humboldt Foundation for partially funding this project.  
Furthermore, we appreciate the assistance of Debashish Mukherji and Raffaello Potestio with computational applications.  
We are also thankful for insightful comments by Tristan Bereau and Robinson Cortes-Huerto regarding the manuscript.  
Finally, we thank the Kavli Institute for Theoretical Physics in triggering this project.

\clearpage
\bibliography{Manuscript}

\begin{thebibliography}{10}

\bibitem{Frisch_ACP1964}
H.~L. Frisch.
\newblock The equation of state of the classical hard sphere fluid.
\newblock {\em Advances in Chemical Physics}, 6(5):229--289, 1964.

\bibitem{Widom_Science1967}
B.~Widom.
\newblock Intermolecular forces and the nature of the liquid state: Liquids
  reflect in their bulk properties the attractions and repulsions of their
  constituent molecules.
\newblock {\em Science}, 157(3787):375--382, 1967.

\bibitem{ChandlerWeeks_PRL1970}
David Chandler and John~D. Weeks.
\newblock Equilibrium structure of simple liquids.
\newblock {\em Physical Review Letters}, 25(3):149--152, 1970.

\bibitem{WeeksAndersen_JCP1971}
John~D. Weeks, David Chandler, and Hans~C. Andersen.
\newblock Role of repulsive forces in determining the equilibrium structure of
  simple liquids.
\newblock {\em The Journal of Chemical Physics}, 54(12):5237--5247, 1971.

\bibitem{ToxvaerdDyre_JCP2011}
Soren Toxvaerd and Jeppe~C. Dyre.
\newblock Role of the first coordination shell in determining the equilibrium
  structure and dynamics of simple liquids.
\newblock {\em The Journal of Chemical Physics}, 135(13):134501, 2011.

\bibitem{IngebrigtsenDyre_PRX2012}
Trond~S. Ingebrigtsen, Thomas~B. Schroder, and Jeppe~C. Dyre.
\newblock What is a simple liquid?
\newblock {\em Physical Review X}, 2(1):011011, 2012.

\bibitem{AndersenWeeks_ACP1976}
Hans~C. Andersen, David Chandler, and John~D. Weeks.
\newblock Roles of repulsive and attractive forces in liquids: The equilibrium
  theory of classical fluids.
\newblock {\em Advances in Chemical Physics}, 34(2):105--156, 1976.

\bibitem{BarkerHenderson_RMP1976}
J.~A. Barker and D.~Henderson.
\newblock What is ``liquid''? understanding the states of matter.
\newblock {\em Reviews of Modern Physics}, 48(4):587--671, 1976.

\bibitem{WeeksBroughton_PRL1995}
John~D. Weeks, Robin L.~B. Selinger, and Jeremy~Q. Broughton.
\newblock Self-consistent treatment of repulsive and attractive forces in
  nonuniform liquids.
\newblock {\em Physical Review Letters}, 75(14):2694--2697, 1995.

\bibitem{PraprotnikKremer_JCP2005}
Matej Praprotnik, Luigi Delle~Site, and Kurt Kremer.
\newblock Adaptive resolution molecular-dynamics simulation: Changing the
  degrees of freedom on the fly.
\newblock {\em The Journal of Chemical Physics}, 123(22):224106, 2005.

\bibitem{Abrams_JCP2005}
Cameron~F. Abrams.
\newblock Concurrent dual-resolution monte carlo simulation of liquid methane.
\newblock {\em The Journal of Chemical Physics}, 123(23):234101, 2005.

\bibitem{EnsingParrinello_JCTC2007}
Bernd Ensing, Steven~O. Nielsen, Preston~B. Moore, Michael~L. Klein, and
  Michele Parrinello.
\newblock Energy conservation in adaptive hybrid atomistic/coarse-grain
  molecular dynamics.
\newblock {\em Journal of Chemical Theory and Computation}, 3(3):1100--1105,
  2007.

\bibitem{PotestioDonadio_PRL2013}
Raffaello Potestio, Sebastian Fritsch, Pep Espanol, Rafael Delgado-Buscalioni,
  Kurt Kremer, Ralf Everaers, and Davide Donadio.
\newblock Hamiltonian adaptive resolution simulation for molecular liquids.
\newblock {\em Physical Review Letters}, 110(10):108301, 2013.

\bibitem{Louis_JPCM2002}
A.~A. Louis.
\newblock Beware of density dependent pair potentials.
\newblock {\em Journal of Physics: Condensed Matter}, 14(40):9187--9206, 2002.

\bibitem{RuhleAndrienko_JCTC2009}
Victor Ruhle, Christoph Junghans, Alexander Lukyanov, Kurt Kremer, and Denis
  Andrienko.
\newblock Versatile object-oriented toolkit for coarse-graining applications.
\newblock {\em Journal of Chemical Theory and Computation}, 5(12):3211--3223,
  2009.

\bibitem{PeterKremer_FD2010}
Christine Peter and Kurt Kremer.
\newblock Multiscale simulation of soft matter systems.
\newblock {\em Faraday Discussions}, 144(0):9--24, 2010.

\bibitem{Noid_JCP2013}
W.~G. Noid.
\newblock Perspective: Coarse-grained models for biomolecular systems.
\newblock {\em The Journal of Chemical Physics}, 139(9):090901, 2013.

\bibitem{ShenHu_JCTC2014}
Lin Shen and Hao Hu.
\newblock Resolution-adapted all-atomic and coarse-grained model for
  biomolecular simulations.
\newblock {\em Journal of Chemical Theory and Computation}, 10(6):2528--2536,
  2014.

\bibitem{HessLindahl_JCTC2008}
Berk Hess, Carsten Kutzner, David van~der Spoel, and Erik Lindahl.
\newblock Gromacs 4: Algorithms for highly efficient, load-balanced, and
  scalable molecular simulation.
\newblock {\em Journal of Chemical Theory and Computation}, 4(3):435--447,
  2008.

\bibitem{BussiParrinello_JCP2007}
Giovanni Bussi, Davide Donadio, and Michele Parrinello.
\newblock Canonical sampling through velocity rescaling.
\newblock {\em The Journal of Chemical Physics}, 126(1):014101, 2007.

\end{thebibliography}

[21] Supplementary Material

\end{document}